\documentclass[a4paper,11pt]{article}

\usepackage{jheppub} 
\usepackage{amsmath,amssymb,amsfonts}
\usepackage{mathtools}
\usepackage{amsfonts}
\usepackage{microtype}
\usepackage{setspace}
\usepackage{lineno,hyperref}
\usepackage{array,multirow,graphicx}
\usepackage{float}
\usepackage[font=scriptsize]{caption}
\usepackage[T1]{fontenc} 

\title{A Double Quantization for 3d Quantum Mechanics with 2d Tiny Extra Window}

\author[a]{Zahra Ghahreman,}
\author[b,1]{Mehdi Dehghani,\note{Corresponding author.}}
\author[a]{and Majid Monemzadeh}

\affiliation[a]{Department of Physics, University of Kashan, 87317-51167, Kashan, Iran}
\affiliation[b]{Department of Physics, Faculty of Science, Shahrekord University, 115, Shahrekord, Iran}

\emailAdd{z.ghahreman@grad.kashanu.ac.ir}
\emailAdd{dehghani@sku.ac.ir}
\emailAdd{monem@kashanu.ac.ir}

\abstract{We construct a quantum mechanics based on the hypothesis of existing compact extra dimensions for a particle that wants to detect it. By introducing a probability function, we express the transition of particle to the extra 2d window. The general properties of this function has been examined and a length scale for occurrence of particle to extra window is given. By a diverse view point we consider that, the new length scale plays another quantum criteria for another quantization, beside the Planck constant. Canonical quantization of second class constrained systems, is our method for constructing the desired quantum mechanics, in which in it the probability function enters in the structure of second class constraints. This import the phenomena of extra dimension to the 3d quantum mechanics, effectively. Some aspects of this effective double quantum theory are mentioned, which one may investigate them more focused to experience extra dimension quantum mechanically in contrast to field theoretic sights. Specially, we try to make solutions for wave function and spectrum of the free particle, by Frobenius prescription for solving linear differential equations. In this context, the length scale of extra dimension characterizes the singularity of the wave equation at the boundary which tiny extra window connected to 3d space. }


\begin{document}
\maketitle
\flushbottom

\section{Introduction}\label{intro}
 Most of the ideas in this work go back to Dirac. We choose a quantum mechanical perspective to the issue of extra dimension  \cite{Arkani-Hamed:1998jmv,ADD}. Our goal is to build a quantum mechanics incorporate to the extra spatial dimensions assumption \cite{Arkani-Hamed:2001kyx,Stojkovic:2013xcj}. The famous prescription for doing quantization on a classical system belong to Dirac. This prescription for an ordinary classical mechanics is straightforward, but Dirac developed it for singular ones which is applied here. Our desired belongs to the second class quantization of Dirac \cite{heno}. We use the theory of constrained systems and its quantization. In this context the Dirac bracket tool helps us to construct desired quantum mechanics.

 We use quantum mechanics for a test particle, as the detector of extra dimensions, because the length of extra dimension occurring forced the particle be tiny. But for simplicity and minimal observations we assume the particle with minimal properties. For example our test particle assumed to has no charges and spin but with  as large as mass in such a way that makes it non relativistic. So we derive a Schr\"{o}dinger equation in contrast to a Dirac equation.

Most models that model particle transitions to extra dimensions, use potentials including the Dirac delta functions, but in our method we introduce functions that causes this transition as an input. In the process of constructing quantum mechanics, we show how Dirac delta functions can
be appear in the problem of extra dimension. But, since it is difficult to work with Dirac delta functions in solving the wave equation, we show how to avoid these functions as more as possible. Difficulty is clear, Dirac delta functions appear in quantum commutator make some functional ambiguity to find representations for the observables. In our method without use the explicit form of the model input, call it $\eta$ in continue, we run away from such difficulties. We define a Planck function $ \hbar(\eta(r))$ by  merging  it in $\hbar$. In this manner by this definition in corporate to cosmic coordinate and cosmic time for the test particle environs the big bang, one can attains another idea of Dirac. It is  the idea of changing the physical constant by the time \cite{Dirac:1974}.

The intention of exploration of extra dimensions by tiny particle was raised for us from a very simple and basic problem in preliminary
 mechanics. Consider a particle on a inclined plane that one going to quantize it. For the object moves on inclined planes, the surface friction coefficient factor can be interpreted as a re-normalized (collected, averaged and coarse grained) factor which is applied to it in dimension less than 3. Such effects keep the particle in the surface and only change its dynamics, effectively. But automatically by some second class constraints the reality of existence extra dimensions could be enters to the problem. In this way, instead of proposing the potential to send the particle to extra dimension, we introduce a distribution function for gaining the coordinate in the extra dimension. The problem must be investigated quantum mechanically, because we consider the test particle to be a small but heavy one. So, we have to make a quantum mechanic for it. Canonical quantization is the best approach to constructing it. New quantum mechanic must be used, because in addition to $ \hbar$ another
  quantum scale, say $\eta$, appears. Also, in constructing process there are some constraints for expressing the existence of extra window.
   The theory of constrained quantization does this. So, in the first part of this paper, with the help of Dirac's instructions, we
   create the reduced phase space of a particle in 3d space with an additional 2d window. There are so many evidences for assuming a
   2d spatial extra window in 4-dimensional space time \cite{Zyla:2020zbs}. From a privative point of view an extra spatial dimension leads to a compactification radios in order of solar system length scale \cite{Rattazzi:2003ea}. The minimal selection for the dimension of extra window is 2 with the compactification radios in a transition domain from millimeter to micrometer \cite{Starkman:2000dy,Giudice:2004mg}. Such domain length for 2d extra window is sufficiently large to acts as a box for a typical massive particle of standard model as a tester of large extra dimension. With these backgrounds we extract the quantum mechanics of a test particle in a (3+2)d background and algebra of its basic operators,  directly.

In the second part, in explaining the particle transition to the extra window, we discuss the particle distribution functions and the possible forms of it, and propose these functions based on the length scale at which the transition to the two-dimensional window occurs. Then we express for them the needed approximations  to solve the relevant Schr\"{o}dinger equation of new quantum mechanic.
In the third section, using the algebra of the basic operators and the possible approximations of the distribution functions, we extract the extra coordinates of the quantum wave equation obtained for the particle and try to solve it by Frobenius method.
Finally in discussion section we propose the left over suggestions about the method of the constructing quantum mechanic and by referring to the derived results suggest some cases for study in detail.
\section{Quantization of 3d particle with 2d window }\label{q3d}
 The problem is to make quantum mechanic in 3d with the assumption that the particle confront to an extra 2d window. With determining the classical phase space we do this aim. The classical detector has the spherical coordinates of a typical three-dimensional space which we add to it a 2d polar coordinate. Based on spherical coordinates, the particle with the functions $ \eta(r,\theta,\varphi)$ and $ \zeta(r,\theta,\varphi)$ obtain coordinates $\acute{\sigma}$ and $\acute{\rho}$ in the extra window. This addition is expressed by following lagrangian constraints, in (3+2)d configuration space.
\begin{equation}\label{1}
    \begin{array}{ll}
       \acute{\sigma}=\eta(r,\theta,\varphi),\;\;\;\;\ & \acute{\rho}=\zeta(r,\theta,\varphi).
     \end{array}
\end{equation}
 The above constrained equations actually select a subspace of the configuration space for particle motion in $\mathbb{R}^5$. The number of constraint equations determines the dimensions of the extra window. In this viewpoint, the particle present in a subset of (3+2)d space, where $\acute{\sigma}$ and $\acute{\rho}$ are the coordinate of particle in the extra window. The quantization will take place in this subspace. Currently, we have considered the types of coordinates $\acute{\sigma}$ and $\acute{\rho}$ to be the same and then converted them to the form of polar coordinates. Other assumption is that the occurrence of our choices are in space and not in time. In this way, the standard form of the classical constraints in the (3+2)d configuration space is as follows.
\begin{equation}\label{2}
    \begin{array}{ll}
       \bar{\phi_{1}}(r;\rho,\sigma)=\rho\cos\sigma-\eta(r), \;\;\;\ & \bar{\phi_{2}}(r;\rho,\sigma)=\rho\sin\sigma-\eta(r).
    \end{array}
\end{equation}
In this notation, in addition to considering the polar coordinates for the extra window we have assumed the space to be homogeneous and isotropic. The homogeneity and isotropy of space has been applied by letting the functions $\eta $ and $\zeta $ as the same functionality and dependency only to the radial coordinate of 3d base space. By a simple technicality in the theory of constrained system \cite{Loran:2002xc}, we can replace the constraints by other linear constraints that are equivalent to them. This is done by solving the constraint equations \eqref{2} mid controlling equivalence conditions and lead to,
\begin{equation}
    \begin{array}{ll}
      \phi_{1}(r;\rho,\sigma)=\rho -\sqrt{2}\eta(r), \;\;\;\ & \phi_{2}(r;\rho,\sigma)=\sigma-\frac{\pi}{4}+2z\pi \;\;\;\ z\in \mathbb{Z},
      \label{3}
    \end{array}
\end{equation}
as new form of constraints. As we have said, these are constraints of the (3+2)d configuration space and are free of any temporal dependencies. Primary constraints may lead to other constraints in the particle's phase space, which can be obtained by examining the free particle dynamics by the following total Hamiltonian
\begin{equation}\label{4}
    \begin{array}{ll}
      H_{T}=H^{(3)}+H^{(2)}+\lambda_{i}\phi_{i}, & \;\;\ \\
      H^{(3)}=\frac{1}{2m}(p_{r}^{2}+\frac{p_{\theta}^{2}}{r^{2}}+\frac{P_{\varphi}^{2}}{r^{2} \sin^{2}\theta}), & H^{(2)}=\frac{1}{2m}(p_{\rho}^{2}+\frac{p_{\sigma}^{2}}{\rho^{2}}).
    \end{array}
\end{equation}
In the total Hamiltonian $H_{T}$, the $\lambda_{i}$ s are Lagrange multipliers that add primary constraints to canonical Hamiltonian.
Apparently in the total Hamiltonian there is no dynamics to send the particle to the extra window. The second term contains only the kinetic
energy of the particle in extra window. But the last two terms in it introduce the interactions that send the particle to
the 2d subspace of the (3+2)d space. The more accurate but trouble maker of the second term can be in the form $\theta(\rho_c-\rho)H^{(2)}$, which by it we have assumed the particle to be in the range $ \rho<\rho_c$.
\footnote{In our model building the  $\rho_{c}$ is a critical length that transition to extra 2d window happened.} In the way \eqref{4} we
avoid the discontinuity of the Dirac delta function appears in our calculations. By such ansatz, we can not discuss about the  detail
dynamics of the transition from 3d space to the 2d window. In fact, by adding primary constraints to the Hamiltonian \eqref{4} we only see transition occur. The primary constraints are added to Hamiltonian by unknown coefficients, where in the formulation of constrained system in order to prevent the more lost of information from the dynamics, we introduce them to the canonical Hamiltonian. At the end of consistency process one can sets the value of constraints zero without anxiety.

 The process of consistency characterize the whole boundary of the phase space in the $\mathbb{R}^{10}$.
So, for the system presented  by \eqref{3} and \eqref{4} the border of the phase space completes by following secondary identities.
\begin{eqnarray}
\bar{\psi_{i}}=\{\bar{\phi_{i}},H_{T}\} \longrightarrow
\begin{cases}
\bar{\psi_{1}}=\frac{1}{m}(p_{\rho}\cos \sigma -\frac{p_{\sigma}}{\rho}\sin \sigma-p_{r}\eta'(r)), \\
\bar{\psi_{2}}=\frac{1}{m}(p_{\rho}\sin \sigma +\frac{p_{\sigma}}{\rho}\cos \sigma-p_{r}\eta'(r)).
\label{5}
\end{cases}
\end{eqnarray}
 It is clear that the particle momentum  in the extra window are related to the coordinates of the base 3d space through the function $ \eta(r)$.
 This issue is specially understood in the total Hamiltonian context that dynamic is examined with all constraints. According to the constraints
  obtained in \eqref{5}, we find that $ H^{(2)}$ acts as part of the potential to send particles to the extra window. It acts in basic
   three-dimensional space provided that total Hamiltonian $H_{T}$ become on shell. It means that we must solve $ p_{\sigma}$ and
   $ p_{\rho}$ from \eqref{5} then replace the $ \rho$ and $ \sigma$ with the help of the constraints obtained in  \eqref{3}. By this procedure we
   keep all the dynamics from lost of information, but do not follow this way for quantization. Instead, we try to identify the phase
   space of the model and build the Hilbert space from it. Therefore, we linearize the constraints obtained in \eqref{5} as much as possible to
    purify them from the contents of the initial constraints obtained in \eqref{3}.
 \begin{equation}\label{6}
    \begin{array}{ll}
      \psi_{1}=p_{\sigma}, \;\;\;\ & \psi_{2}=p_{\rho}-\sqrt{2}\eta'(r)p_{r}.
    \end{array}
\end{equation}
Now the set of constraints obtained in \eqref{3} and \eqref{6} are second class. They introduces the phase subspace $ \mathcal{M}_{6}\subset\ \mathbb{R}^{10} $ as the available phase space for the classical model. The matrix of Poisson bracket between constraints ( $\Delta$ in below)  gives us the structure of reduced phase space.
\begin{eqnarray}
&&\Omega_{IJ}^{(\eta)}=\{z_{I},z_{J}\}-\{z_{I},\psi_{\alpha}\}\Delta_{\alpha\beta}^{-1}\{\psi_{\beta},z_{J}\}, \nonumber \\
&& z_{I}\in\{r,\theta,\varphi,\rho,\sigma,p_{r},p_{\theta},p_{\varphi},p_{\rho},p_{\sigma}\},\nonumber \\
&&\psi_{\alpha}\in\{\phi_{i},\psi_{i}| i=1,2\}.
\label{7}
\end{eqnarray}
 In the above structure construction, the components of the symplectic two form $\Omega$ is calculated by famous Dirac formula. The set
  $(\mathcal{M}_{6},\Omega^{(\eta)})$ shows that the assumption of the presence of extra dimensions causes the immersion of manifold $
  \mathcal{M}_{6}$ in $ \mathbb{R}^{10}$ with a symplectic structure that is controlled by $\eta(r)$. This form of introducing phase space for
   Poisson manifold, which corresponds to systems with finite degrees of freedom, can be explained in the context of Kontsevich's quantization
   method \cite{Kontsevich:1997vb}. The 6-dimensional manifold of the reduced phase space indicates that the particle is still in the 3d
   configuration space, but the only effect of the extra dimensions is quantization deformation. Deformation quantization method for it, is more
    interested if one pays attention to the fact that primary manifold of configuration space is in the form of attaching a cell in the form of
     ball (2d window) to the $\mathbb{R}^3$ by gluing process, where the \eqref{3} play as attaching map \cite{yenishbook}. From quantum cosmological aspects for such attaching and gluing 2d window to 3d space, we can notice the \cite{Carlip:2017eud,Afshordi:2014cia}.

  For the basic variables of the model, the Poisson structure obtained from \eqref{7} after canonical quantization is obtained as the following commutator for the 3d part of phase space with oneself basic objects.
\begin{eqnarray}
&&[r,p_{r}]=\imath\hbar(r),  \hspace{0.3 cm}   [\theta,p_{\theta}]=\imath\hbar,  \hspace{0.3 cm} [\varphi,p_{\varphi}]=\imath\hbar.
\label{8a}
\end{eqnarray}
In which, due to the presence of extra dimensions controlled by function $\eta(r)$, the Planck constant is transmuted into a heterogeneous and isotropic Planck function,
\begin{eqnarray}
&& \hbar(r)=\frac{\hbar}{1+\eta'^{2}(r)}.
\label{9}
\end{eqnarray}
The heterogeneity of the Planck function is seen in its dependence on $r$. Dynamical Planck quantities, was investigated by the notion of Planckion scalar in the realm of effective field theory around the TeV scale, via the data from BICEPs, KEK and PLANCK collaboration \cite{Kannike:2015apa}. So, one can relates $\eta $ to the wave function of Planckion in the first quantization of a quantum gravity as a field theory at big bang epoch and choose the new physics of \cite{Kannike:2015apa,Kannike:2016wuy} as standard model with an extra window.

Quantum commutators between basic observables of 3d and 2d parts characterized by $\hbar$ and $\eta$, as fallow
\begin{eqnarray}
&& [p_{r},\rho]=\imath\eta'(r)\hbar(r), \hspace{0.3 cm}    [p_{r},p_{\rho}]=-\imath\hbar(r)\eta''(r)p_{r}+O(\hbar^{2}), \hspace{0.3 cm} [p_{r},p_{\sigma}]=-\frac{2\imath \hbar(r)\eta'(r)}{\rho}p_{\sigma}, \nonumber  \\
&& [p_{\rho},\rho]=-\imath\hbar(r)\eta'^{2}(r), \hspace{0.3 cm}   [p_{\rho},p_{\sigma}]=-\frac{2\imath \hbar(r)\eta'(r)}{\rho}p_{\sigma}, \hspace{0.3 cm} [p_{\rho},r]=\imath \hbar(r)\eta'(r).
\label{8}
\end{eqnarray}
The phrase double quantization refers to the quantization by $\hbar$ and $\eta$ and becomes clear from above basic commutators.
In decade 1930 and afterward 1970, the temporal heterogeneity of fundamental physical constants was proposed by Dirac \cite{Dirac:1974} named large numbers hypothesis. This idea is a candidate for solving dark energy and dark  matter problems. In some seance this idea can we expanded by the founded commutators, because the problem belong to the territory of quantum cosmology. At the cosmic time $t_{\textit{cosmos}}$, Planck's constant \eqref{9} heterogeneity can lead to its temporal heterogeneity. Therefore the mathematical formalism of Dirac's constrained systems for assuming extra dimensions leads to other idea for investigating the problem of matter and dark energy, $ \hbar(t_{\textit{cosmos}})=\hbar(r(t_{\textit{cosmos}}))$.

Before investigating the quantum wave equation in the model, it is necessary to point out the quantum anomaly that has appeared in construction above quantum commutators. In the set of quantum commutators, the commutators between radial momentum has an specials form. According to it and by symetrization of the multiplication between non-commutating  operators, we find,
\begin{eqnarray}
&& [p_{r},p_{\rho}]=-\imath\hbar(r)\eta''(r)p_{r}+\frac{1}{2}\hbar^{2}(r)(\eta'''(r)-\frac{2\eta'(r)\eta''^{2}(r)}{1+\eta'^{2}(r)}).
\label{10}
\end{eqnarray}
The first term in the R.H.S is an ordinary canonical quantization effect. But the second term is of order $ \hbar^{2}$, so is a quantum anomaly
which can be studied  only in the interfering patterns. Such pure quantum effect for the test particle can be revealed via  quantum phases,
geometrical Berry phase \cite{nakahara}, in areas containing all 5 coordinates. By such interpretations the result \eqref{10} means that, if part
of the path traveled by the test particle  through an extra 2d window, then an extra topological phase  will occur which is a sign of extra
dimensions. More investigation of such effects can be down in the realm of quantum field theory, because it's order of magnitudes belongs to second quantized system. Nevertheless, if in quantum mechanic scope consider that the number of dimensions is a topological feature, then changing 3d to
 5d has changed the topology of the space. So, when we assemble a geometry on a topological space by a length parameter ($\rho_{c}$), the
  expectation for the emergence of topological objects \cite{Hill:2001bt} satisfied but their interactions \cite{Bai:2009ij} is the lack of
  quantum mechanical model building in contrast to field theoretical approach.\\
  \;\;\;\;\ In addition to above mentioned Berry phase as topological object, suppose we want to write the fundamental constant of a Dirac magnetic monopole. For this purpose using fundamental constants of the 3d universe and with the fundamental constants attributed to the added 2d window, say $\rho_{c}$, inevitably we lead to the fundamental electric charge $e$ and a fundamental mass $ \textit{m}$ appear in it.
\begin{eqnarray}
&& \mathbf{B}=\frac{Q_m}{r^{3}}\mathbf{r}, \;\;\; Q_m=\frac{\textit{m} \rho_{c}}{e}c.
\end{eqnarray}
In this manner, the Dirac quantization condition for the magnetic charge $Q_m$ leads to the quantization of mass.
\begin{eqnarray}
&& \textit{m}=\frac{1}{2}\textit{z}\frac{\hbar}{\rho_{c}}, \hspace{1 cm} z \in \mathbb{Z}
\label{10a}
\end{eqnarray}
That is, we can achieve to the quantization of mass and even so the agravity \cite{Salvio:2014soa} by combining the assumptions of the existence of extra dimensions $(\rho_{c})$ and magnetic monopole. The mass scale derived here( \eqref{10a}) is tuned by the quantum constants of the two parts, $ \hbar$ belongs for 3d part and $(\rho_{c})$ for 2d extra window. Two interesting value for $(\rho_{c}) $ is that, if we choose the size of the extra dimension of the order of millimeters, the quanta of mass becomes of the order of the electron mass and by the Fermi order for $\rho_{c}$ we obtained a larger quanta  for the mass, say $10^{-19}$kg or TeV scale. Where the last one recently show to concluded from compact extra dimension by Poincar\'{e} symmetry considerations  \cite{Mathieu:2020ywc}. Therefore it is shown that extra 2d window incorporate to the Dirac magnetic monopole assumption can be resulted to the negative gravity and mass quantization.
\section{Toward the particle spectrum}\label{toward}
In the continuation of our study, we want to obtain the particles energy spectrum  up to the approximation of the lowest order of $ \hbar \eta$. Since the anomalous  term in the \eqref{10} is inherently of the order of $\hbar^{2}\eta$, the resulting corrections from it is automatically excluded. Moreover, the presence of this term makes the representation of the fundamental observable too problematic, so its omitting is welcome.
Deriving the basic commutator are the first step in constructing any quantum mechanics. The next step is to obtain the Hilbert space of the
particle in this quantum mechanics. For the quantum mechanics based on minimal length consideration, which according to $\rho_c$ the present model fall in its categories, both steps was done in \cite{Kempf:1994su}, so is a good hint here. 

For our case in order to find the eigen states we have to find the representation of the basic
operators from their algebra \eqref{8} and \eqref{8a}. Basic commutator in \eqref{8a} suggests that position state is still the best eigen state for representations. Therefore, the change in the representation of operator $ p_{r}$ is due to change in radial commutator. 
This happened because the particle is transferred to the extra window by it. According  to the isotropic assumption, the other two momentum operators don't transfer the particle to the extra window, so they give their usual form $(\hat{p_{\theta}}=-\imath\hbar\partial_{\theta},\hat{p_{\varphi}}=-\imath\hbar\partial_{\varphi})$.

 In order to find the modified representation of $p_{r}$, we have to pay attention to the fact that, this operator also makes commutator with other operators. So its proposed representation must be compatible with other commutator, as well. The most complex commutator is $[p_{r},p_{\sigma}]$. However, it does not cause any trouble or limitation to finding the representation for $ p_{r}$, because due to the quantum version of constraint \eqref{6}, $\hat{p_{\sigma}}$ is a null operator. It means that the momentum acting in the extra window is not physically observable in 3d part. It is a trivial topological value for 3d observers. Before finding the physical phase space note that $\hat{\sigma}$ has same behavior, i.e. after finding the reduced phase space and quantizing it acts as a c-number.
\begin{eqnarray}
&&\hat{\sigma}=\frac{\pi}{4}+2\pi z , \hspace{1 cm}z\in \mathbb{Z}.
\label{11}
\end{eqnarray}
Although, this parameter doesn't applicable in our model because we are examining the effects of the extra window outside the window, the effect of \eqref{11} parameter may appear in the detection of anomaly \eqref{10} by interferometry. Also, the integer $z$ in this formula refer to a topological quantum number as winding number for test particle around and into the extra window \cite{nakahara}.

Conclusively the null representation of  $\hat{p}_{\sigma} $, removes two of the messy commutator in relations  \eqref{8}, leads us to following  representation :
\begin{eqnarray}
&&\hat{p}_{r}=-\imath\hbar(r)\partial_{r}, \hspace{1 cm} \hat{p}_{\rho}=-\imath\hbar(r)\partial_{\rho}, \hspace{1 cm} \hat{\rho}=\eta(r).
\label{12}
\end{eqnarray}
Now, with the help of the representations that we found for the observables, we know that all the calculations and the Schr\"{o}dinger  equation are in the  3d part coordinate. We don't need to justify how we choose a potential to send the particle to the 2d window. Such potentials can be obtained from  $ \eta(r)$. Also, with this reduction in the classical form, we can calculate the gravitational potential corrections due to the presence of extra window. However, the explanation of how to select function $ \eta(r)$ will be the same as the explanation of how to select metric and potential to sending particle to extra dimension. 
\section{Wave equation in 3d}\label{wave3d}
We are now in the position to render the wave equation of the model according to the Hamiltonian \eqref{4} and operator
representation \eqref{12}, as a partial differential equation. We only work with canonical part of Hamiltonian because the effects
 of second class constrained have arrived by Dirac bracket, conclusively one can let constraint vanishing strong. In this way the full effects of constraints and $H^{(2)}$ part of Hamiltonian play the role of some  double quantum potential.\footnote{double quantum potential means that some parts belong to $\eta$ and some parts to $\hbar$.} The Schr\"{o}dinger  equation illustrates this fact.
\begin{eqnarray}
-\frac{1}{2m}[\hbar \hbar(r)\partial^{2}_{r}+(-\eta'(r)\eta''(r) \hbar^{2}(r)+\frac{2}{r} \hbar\hbar(r))\partial_{r} -\frac{1}{r^{2}}\mathbf{L}^2] \psi(r,\theta,\varphi)=E\psi(r,\theta,\varphi)
\label{13}
\end{eqnarray}
This equation shows that the metric of the (3+2)d space appears as $1+ \eta'^{2}(r)$ in the appropriate positions, eke the potential of sending the particle to extra window is given in Schr\"{o}dinger equation with derivatives of $\eta(r)$. In this case, due to the homogeneity of the occurrence of extra dimensions, $ \mathbf{L}^2$ separated from the radial part directly. In fact, the change in the special Laplace operator of the builded model due to the change in the metric, only appears in radial part. Also the potential of sending the particle to the extra window is radial as well as momentum. Eventually, the standard form of the radial equation after inserting the quantum number of angular momentum is as follows.
\begin{eqnarray}
&& R''(r)+P(r)R'(r)+Q(r)R(r)=0\nonumber \\
&& P(r)=\frac{\hbar'(r)}{2\hbar(r)}+\frac{2}{r}\nonumber \\
&& Q(r)=-(\frac{l(l+1)\hbar^{2}}{r^{2}}-2mE)\frac{1}{\hbar\hbar(r)}
\label{14}
\end{eqnarray}
In the following we intend to solve Equation \eqref{14} for the eigen state energy. Therefore we need to know the explicit form or general behavior of $\eta(r)$. In the next section we introduce this function by its properties. Thence, after identifying its general behavior, we proceed to solve Equation \eqref{14}.
\section{Occurrence in extra window }\label{occore}
In line with the attempt to find solutions in constructed doubled quantum mechanics, here we try for wave function in the 2d window part. The spirit of the content presented here are like the ones people offered in the beginning of quantum mechanics, at times when still didn't have wave equation.

At the beginning \eqref{1} we consider general simplifying assumptions for this function. In the following we present other assumptions according to our expectations of this function. In abecedarian, let us state that the physical unit of this function is of the form of length. As we discussed after \eqref{12}, this function can  play two role, a metric role and the role of potential to sending quantum object to the extra window. In another viewpoint this is a distribution function which around a point gives the probability for finding a extra window. In other word,  it says how a particle selects the coordinates of a 2d extra window $(\rho,\sigma)$ around a general point in 3d space. 
It is clear that such a path selection depends on both the 3d space coordinates and the extra window coordinates, 
 i.e  $\eta( \mathbf{r},\mathbf{\rho})=\eta_{1}(\mathbf{\rho})\eta_{2}(\mathbf{r})$. 
 Of these two $(\eta_{1}(\mathbf{\rho}),\eta_{2}(\mathbf{r}))$, the  observer senses the coordinates of the 3d space. 
 The maximum length that the observer observes in the extra window is the critical threshold length, say $ \rho_{c}$. At this scale, 
 the particle passes in and out the extra window. Therefore, in the coordinate system of this observer, $\eta (\mathbf{r},\mathbf{\rho})$ is written as $\eta(\mathbf{r},\rho_{c})$. The parameter $ \rho_{c}$ which characterizes the model, is the upper limit for the value of  extra window say $\rho$. So, if $r>\rho_{c}$ we don't see the occurrence of extra dimensions. In other words if the energy level of the particle is higher than the value specified by $\eta(r,\rho_{c})$, an extra window will open for the particle. According to \eqref{3}, the extra window surface is proportional to $\eta^{2}$. This is the scattering cross section of the particle into the extra window. For this reason, we interpret $\eta$ as a probability distribution function.

Although we don't know anything about the inside of the extra window and its attractive potential to absorb particles, but a choice for the inside of the  extra window could be as follows,
\begin{eqnarray}
&& \eta_{1}(\rho)=\rho^{2}+ \alpha \rho_{c}\rho+\beta \rho_{c}^{2}.
\label{15}
\end{eqnarray}
Where this suggestion is based on dimensional analysis with $\alpha$ and $\beta$ as dimensionless geometrical numbers.
The first term and the larger powers of the $\rho$ should not be present in the above relation because in this case at any energy, the particle may be transits into the extra window, nowhere such a phenomenon that has been observed or it refer to instability. The second term disturbs the homogeneity of the space inside the extra window, therefore we leave it aside. Thus the first choice naturally is $ \eta_{1}(\rho)\sim \rho_{c}^{2}$.

 For the part $ \eta_{2}(r)$, there are many choices that can be better verified. The first expected feature for these choices is that they are oscillating in radial coordinate. If the space is homogeneous for the occurrence of extra windows, the wavelength of these oscillations, which can be controlled by $\rho_{c}$, must be very small. Of course, this homogeneity of space must be in both 2d and 3d parts. For example if the 2d part not be a homogeneous one, then the particle scattered into the extra window will not return to the original 3d space. The second expected feature for $\eta$ is that it must be damped at a specific  value of $ r$. By mathematical tools it means that the distribution function around the specific point in the limit $ \epsilon=\frac{\rho_{c}}{r}\rightarrow 0$, approaches to zero. Therefore, this function requires a damping section in addition to it's oscillating section. In the absence of a damping section, the particle even at low energies, may accidentally gains a significant portion of $ \eta$ and enter the extra window.
 
So far we have said two main points about $\eta(r)$. First, such functions consists of two parts, in which we used a critical length to write these parts. Second, critical length is a boundary for $ \eta(r)$. We must specify the boundary conditions for it. If we consider a specific point and its neighborhood with a radius of $ \rho_{c}$ , its infinite boundary conditions have already been expressed by the appropriate variable $\epsilon=\frac{\rho_{c}}{r}$. In the case of boundary $ \rho_{c}$, if we want to start with an ideal selection for $ \eta(r)$, we must take a step function at this point. In this case the derivative of $\eta(r)$ is a Dirac delta function appears in quantum commutators. To fix this difficulty  and soften the stair-like fracture of the extra window edge, $ \rho_{c}$ acts as an ultraviolet cut-off. On the other hand, choosing a specific physical item that wants to experience the phenomenon of extra dimensions, such as a particle in a box, gives us an infrared cut-off. In the case of the particle in the box, this infrared cut-off is the length of the box $L$ that gives us $ \epsilon=\frac{\rho_{c}}{L}$. So, in the case of a free particle $\epsilon=0$. Conclusively, for the wave function of the particle inside the extra window we can't get it in the form of a convergent series. But, inside the extra window doesn't matter to us. We do our calculations outside of the window and want to discover the effect of the presence of extra dimensions in this way.

After the above explanation, we will examine the possible forms for function $ \eta(r)$. We found that the most ideal choice for $ \eta(r)$ is a constant function, say a periodic function with zero wavelength that turned on in  $ 0<r<\rho_{c}$ and turned off outside this area. Also it must be homogeneously and isotropically repeated in different directions. Such a function $ \eta(r)$, its derivative and its linear combination, produces the periodic Dirac delta  function in space as $\sum _{n}\delta(r-n\rho_{c})$, which is reminiscent for the potential of the Kronig-Penney model in quantum mechanics. This model was use for repulsive hard core dots. In corporative to first derivative, the second derivative of  $ \eta(r)$ is also included in both commutator and potential, so in calculations. This makes it even harder to work with this ideal form of $ \eta(r)$ but also harder than Kronig-Penney model. The first trick to solve these problems is to use Delta function representations. Of course, it complicates the representation of operators and wave equations and take them out of the standard form of a differential equation. Using the Gaussian display for the Delta function (finally one can tend its width to zero) is another way to solve this problem. But hit-or-miss where we need to expand it, it has difficulty converging. We may think that the standard form of the Gaussian function is the cause of these problems. But even  we consider the following special form, in which the oscillating and damping parts are separated, then we find that we still have a fundamental singularity at $\epsilon=0$.
\begin{eqnarray}
&& \eta(\epsilon)=\rho_{c}\epsilon \exp(-\beta\epsilon)\sin(\frac{\alpha}{\epsilon})
\label{16}
\end{eqnarray}
The parameters $ \alpha $ and $ \beta$ regulate the attenuation and wavelength of the $ \eta$. Both can be introduced in terms of $ \rho_{c}$. They themselves are connected with the help of the normalization condition introduced for $ \eta$. The attenuation wavelength and the wavelength of $ \eta$ are related to each other as seesaw, i.e any value for one of the to smooth  $ \eta$ on one boundary affects the other boundary. These are signs of the phenomenon of mixing IR and UV in which a minimal length (here $ \rho_{c}$)cause it.
Equation  \eqref{16} shows that if there exists an expansion for  $\eta$, it is a Laurent series in contrast to the Taylor expansion. In this way depending on the physical purpose of the problem one  decide to keep his (her) desired terms of expansion. This means that from the 3D observer's point of view there is fundamental singularity in $ \eta$, but fortunately we don't need wave function near the singularity, since we do not want to use observables inside the extra window.

 As a conclusion of the discussion in this election of $ \eta$, one feels that we are talking about something like the basic quantum mechanical wave function. This function specifies the distribution of space bumpiness, holes and extra 2d windows, for the particle. When we find this function, we put it in the quantum wave equation \eqref{13} obtained to investigate the behavior of the test particle in this extra window. This means that we find another wave function in 3d space. The $ \eta$ in addition to having a role in determining the metrics and coordinates of space, it also has a role as a potential for test particles. Although we don't know the wave equation for it but we assume that there is a duality between it and the particle wave function in 3d, say $\psi(\mathbf{r},\theta,\varphi)$. In other words, if $ \eta$ be a Planck function in the Schr\"{o}dinger wave equation, there is a duality between the wave function and the Planck function. In a way, $\psi(\mathbf{r})$ is a functional of the Planck function and vice versa.
\begin{eqnarray}
&& \psi=\psi(\mathbf{r};\eta(\mathbf{\rho})]\leftrightarrow \eta(\mathbf{\rho};\psi(\mathbf{r})]
\end{eqnarray}
Because two parts of the whole 5d wave function relate to each other by the conjectured duality, it is belong in the type of coordinate-wave function duality of Faraggi-Matone \cite{Faraggi:1996rn}. Dualities in and between the theories have the topological aspects.  The anomaly we obtained in the previous sections, which was controlled by $ \rho_{c}$, could confirm the topological object belong to this duality, something like Dirac magnetic monopole. However, this survey needs further study, which we will discuss later. But what we learn from this fact swiftly, is that the border of 2d window and 3d universe is not like an event horizon of a black hole. According to the general properties of $ \eta$, we solve the wave equation in the 3d space for a particular case to obtain the local results in contrast to topological ones. Equations and their solutions also may investigated deeply to give us more information about the inside of 2d window. So In subsequent calculations, although we have entered $ \eta$ with its principal properties in the Schr\"{o}dinger equation, only a small fraction of the singularities in the wave equation go back to function $ \eta$, and the equation itself is the principal cause of the singularities. 
\section{Frobenius method for wave function}\label{frob}
With the help of model's observables in 3d sub-space from whole space, we obtained quantum wave equation. In particular, we represent the $\eta(r)$ by the observables of the 3d base space. Now it is the time to extract the corrections on the wave function that come from the $\eta(r)$. The simplest case for study in 3d space, is the free particle, as well as for builded our model. The absence of the external source for potential allows one to does the process of separation of the equation to be performed in the kinetic part. In fact, must of the calculation is separation in the (3+2)d Laplacian. The 2d part consists of an angular part which is a gauge constant according to \eqref{11} and a radial part. The radial part was related to the radial coordinate of the 3d space with the help of constraint equations \eqref{3}. In this way, the (3+2)d Laplacian corrections are of the radial type and its angular part is separated by solving spherical harmonics. The parts of the radial operator that contain $\eta(r)$ are interpreted as the model's internal potential for sending the particle to the extra window.

Although, in the following we try to solve the equation to find the wave function, but most of our attention is to find eigenvalues of energy that corrected by $\rho_{c}$. As mentioned before, in 3d space and in the region $ \rho_{c}\leq r<\infty$, we solve the radial equation \eqref{14} by Frobenius method with the dimensionless variable $\epsilon=\frac{\rho_{c}}{r}$. The dimensionless variable and functions help us to choose suitable terms of expansion of the series solution for a converge answer. The power series expansion for $\bar{P}(\epsilon)$ and $\bar{Q}(\epsilon)$ originates  from the fact that we can write $ \eta(\epsilon)$  as a polynomial according to $\epsilon$.
the definition of $\epsilon$ as the inverse of radial distance refer to the this fact that we want to solve the problem from the point of view of a 3d observer of 5d space or even another observer in a 2d window. Accordingly, the $ r=0$ is not the point under consideration for 3d observer.

Based to the above explanations, the radial equation transforms to,
\begin{eqnarray}
&& \bar{R}''(\epsilon)+\bar{P}(\epsilon)\bar{R}'(\epsilon)+\bar{Q}(\epsilon)\bar{R}(\epsilon)=0.
\label{18}
\end{eqnarray}
 In this equation the desired function $\bar{R}(\epsilon)$ and apparent function ($\bar{P}(\epsilon)$ , $\bar{Q}(\epsilon)$ )are obtained by redefining and dimensioning the functions ( $R(r)$, $P(r)$, $Q(r)$) in the initial radial equation \eqref{14}, as follows.
\begin{eqnarray}
&& \bar{P}(\epsilon)=-\frac{\epsilon^{3}\bar{\eta'}(\epsilon)(2 \bar{\eta'}(\epsilon)+ \epsilon \bar{\eta''}(\epsilon))}{1+\epsilon^{4}\bar{\eta'}^{2}(\epsilon)},\nonumber \\
&& \bar{Q}(\epsilon)=-\frac{1}{\epsilon^{4}}(\epsilon^{2}l(l+1)-\bar{E})(1+\epsilon^{4}\bar{\eta'}^{2}(\epsilon)),\nonumber \\
&&\bar{R}(\epsilon)= \frac{1}{\rho_{c}^{2}}R(\epsilon).
\label{19}
\end{eqnarray}
Where the dimensionless quantities of equation \eqref{19} are defined as follows:
\begin{eqnarray}
&& \bar{\eta}(\epsilon)=\frac{1}{\rho_{c}}\eta(\epsilon)  \hspace{2 cm} \bar{E}=\frac{2mE}{\hbar^{2}}\rho_{c}^{2}
\label{20}
\end{eqnarray}
In the following the smallest approximations, say up to second, for the $\bar{\eta}(\epsilon)$ can be sufficient for our single scale
calculations. We are more inclined to find the energy spectrum and more specifically to correct the ground state energy to show what is the effect of the extra dimensions in the lowest states. It is clear that  $\epsilon_{0}=0$ is an irregular singularity for equation \eqref{18} because the fires criterion in the Frobenius method in following list is finite but the second one is not. As a remainder notice that the list of criteria is
\begin{eqnarray}
&& \lim _{\epsilon\rightarrow \epsilon_{i}}(\epsilon-\epsilon_{i})\bar{P}(\epsilon)<\infty, \hspace{2 cm} \lim _{\epsilon\rightarrow \epsilon_{i}}(\epsilon-\epsilon_{i})^{2}\bar{Q}(\epsilon)<\infty,
\label{21}
\end{eqnarray}
where $\epsilon_{i}$ is a singularity point of a second order differential equation. The emergence of the irregular singularity is due to the 
global effects of the existence of extra window. It means that at $\epsilon\rightarrow 0$ we may have avoided the effects of divergence due to
 the presence of an extra window around $r$, but we have inevitably entered the windows around adjacent points  and this causes an uncontrollable divergence in  $\epsilon_{0}=0$. But for  $\epsilon_{1}=1$, both criteria \eqref{21} have finite value, which means that this point is regular. 
 In fact, if we examine the model locally, we do not encounter  mathematical hardship.
Now, with the launch of the Frobenius machinery, we start solving \eqref{18} around $\epsilon_{1}$.
By putting  the series $ \bar{R}(\epsilon)=\sum_{\lambda=0}^{\infty}a_{\lambda}(\epsilon-1)^{\lambda+k}$ in the wave equation, we get 
the characteristic equation for $k$. It determines the degree of allowable divergence for the answer. Any value for $k$ obtained from the characteristic equation, gives us an allowable answer for the wave function. Each  $\bar{R}(\epsilon)$ also has its own energy. By determining 
the energy of each wave function and sort them, the energy spectrum is obtained. The ground state energy of a typical free particle is zero. By calculating the energy for $ k=0 $ and $ k=1$, it is clarified that the ground state energy in this model arises from $k=0$. In zero order approximations for $\bar{P}(\epsilon)$,  $\bar{Q}(\epsilon)$ and for $k=0$, we find the following equations,
\begin{eqnarray}
&& \bar{q_{0}}=\bar{Q}(1),\;\; \bar{q_{0}}a_{0}=0,
\label{22}
\end{eqnarray}
where( $\bar{p}_{n}$ , $\bar{q}_{n}$) are the Taylor expansion of the coefficients of \eqref{19} around $\epsilon_{1}$. By solving equations \eqref{22} and considering non-trivial wave functions, the following equations for energy and $\eta$ are obtained.
\begin{eqnarray}
&& \bar{\eta}(1)=\exp(\imath\pi(z+\frac{1}{2}))  \hspace{1 cm}z\in \mathbb{Z},\nonumber \\
&&E_{(l,\eta)}^{0,0}=\frac{l(l+1)\hbar^{2}}{2 m \rho_{c}^{2}},
\label{23}
\end{eqnarray}
here $l$ is the usual quantum number of 3d space angular momentum. Also the whole spectrum of the particle at this stage is unknown but these two answers, sounds to a discrete spectrum in the continuum of the energy of the free particle. But the continuous part, that results from the first part of \eqref{23}, is unacceptable and must be omitted because it causes divergence in $\epsilon_{1}$ and make ambiguity in the representation of the 3d operators \eqref{12}. However, subtilizing to the fact that the excluded part of the answer, provides information about the inside of the extra window, is the particle wave function inside the window, i.e $\eta$). Attempting for the following determinants equation, guided us to other parts of this spectrum for higher order approximations around $\epsilon_{1}$.
\begin{eqnarray}
\begin{vmatrix}
 \overline{q}_{0} & \overline{p}_{0} & 2 & 0 &0  &\ldots\\
\overline{q}_{1}& \overline{p}_{1}+\overline{q}_{0} & 2\overline{p}_{0} & 6&0 &\ldots\\
\vdots & \vdots &\vdots & \vdots\\
 \overline{q}_{n} & \overline{p}_{n}+\overline{q}_{n-1} & 2\overline{p}_{n}+\overline{q}_{n-2} & 3\overline{p}_{n-2}+\overline{q}_{n-3}&\ldots \\
\end{vmatrix}
=0
\label{24}
\end{eqnarray}
The another part of the spectrum is similarly derived by solving the corresponding equations for the other root $(k=1)$ of the characteristic equation. For example, the energy of the first excited state for $k=1$ is obtained as follows.
\begin{eqnarray}
&& E_{(l,\eta)}^{(1,1)}=\frac{\hbar^{2}}{2 m \rho_{c}^{2}}[l(l+1)-\frac{1}{1+\eta'^{2}}(\bar{P}(\epsilon)+ P'(\epsilon)]\biggr\vert_{\epsilon=1}.
\label{25}
\end{eqnarray}
In \eqref{24} and \eqref{25}, the terms with quantum numbers $l$, indicate that the ground state of the particle in the 3d part of space has zero energy($l=0$). The first part in these spectra, resulted from  $k=0,1$, can be refers to a rotating term with rotational inertia $m \rho_{c}^{2}$ in the effective Hamiltonian for particle living in this model. The phrases $\bar{q}_{n}$ in determinant \eqref{24}
\begin{eqnarray}
&& \bar{q}_{n}=\kappa^{(n)}(1)(l(l+1)- \bar{E})+n l(l+1)(\kappa^{(n-1)}(1)+(n-1)\kappa^{(n-2)}(1))\nonumber \\
&& \kappa(\epsilon)= -\frac{1+ \epsilon^{4}\bar{\eta}'^{2}(\epsilon)}{\epsilon^{4}}
\label{26}
\end{eqnarray}
and other expressions related to the calculation of the particle energy spectrum, show that such a term ($ \frac{\hbar^{2}l(l+1)}{2 m \rho_{c}^{2}} $) always exists in particle energy spectrum. This term refers to the roaming the particle around the soft core of the extra window. In such a way that the particle may passes through or scatters into the soft window  with the probability calculated by $\eta(\epsilon)$. Therefore, its energy spectrum is generally as follows,
\begin{eqnarray}
&& E_{(l,\eta)}^{(k,n)}=\frac{\hbar^{2}}{2 m \rho_{c}^{2}}[l^{2}+l +f_{l}[\eta(\epsilon),\eta^{(1)}(\epsilon),\cdots,\eta^{(n+3)}(\epsilon)].
\label{27}
\end{eqnarray}
The first term $\frac{\hbar^{2}l^{2}}{2 m \rho_{c}^{2}}$, which always appears in the energy spectrum that clearly independent
of the model characteristic function($ \eta(\epsilon)$), refers to Kaluza-Klein modes. Strictly speaking it can refer to gravy scalars
of \cite{Giudice:1998ck}. Examination of \eqref{27} with the reference \cite{Hannestad:2003yd} confirmed that the dimensionality of
extra window is 2 and $ \rho_c\sim 1.6\times 10^{-10}m$. It is worth to nothing that such gravy scalar bounds come from astrophysics evidence
 specially from neutron star and SN1987A data, but the aforementioned $ \rho_{c}=0.5 mm $ belongs to internal consistency of extra dimensions models.

The second term inherits the  quantum oscillation of the particle on the boundary $\epsilon_{1}$, with $ \omega=\frac{\hbar}{2 m \rho_{c}^{2}}$. Therefore, the Casimir force resulting from the assumption of extra dimensions can be derived as $|F|\sim\frac{\hbar}{\rho_{c}^{3}}$. In Casimir energy view point to extra dimension, an amount for length of window derived $35\mu$m by astrophysical data \cite{Dupays:2013nm} which one could uses to bring his(her) to the true form of $\eta$ from those suggested in \eqref{15} and \eqref{16}, phenomenologicaly.

The last term in the energy spectrum is related to the choice of $ \eta(\epsilon)$ or details of the model. There is no guarantee that this term be necessarily real numbers or even become positive real numbers by collaborating on the previous two terms. This means that positive and negative real numbers or even complex numbers can be present in the energy spectrum. So here, appearance for complex energy means that the particle transits into the window  and return from it. The imaginary part in a typical complex energy means the  transition width, so from the imaginary part one can calculates the transition time to the extra window. Thus, the boundary $\epsilon_{1}$ is the surface of the Dirac sea, that the 3d space is above it and the 2d window is below it. The emergence of the complex energy with negative real part can be interpreted as creation of a hole below the Dirac sea, so an anti-particle in 3 space.

Finally,  let us pay attention for the characteristic $k=1$. It's equations like  \eqref{24}, don't form a set of $n$ linear equations with $n$ variables. In corporate to equations like\eqref{24} for spectrum, we also get equations for $ \eta(\epsilon)$. For example, in the second order approximation, we get the following three equations for the 3d wave function.
\begin{align}
& \begin{cases}
2 \overline{a}_{1}+ \overline{p}_{0}a_{0}=0\\
2\overline{p}_{0}a_{1}+\overline{p}_{1}a_{0}+\overline{q}_{0}a_{0}=0 \\
a_{0}\overline{p}_{2}+2 a_{1}\overline{p}_{1}+\overline{q}_{0}a_{1}+\overline{q}_{1}a_{0}=0.
\end{cases}
\end{align}
\label{28}
Solving the set of above equations for $a_{i}$s, surprisingly  leads to a set of conditions for the wave function inside the window and the
 particle energy spectrum in 3d. The conditions obtained for the wave function inside the window are expressions for derivatives of
  function $\eta$ on the boundary $\epsilon_{1}$. The greater number of equations like \eqref{24}, leads to the more boundary conditions 
  for $\eta$. So, it gives us a more accurate Taylor expansion for $\eta$ at point $\epsilon_{1}$. This means that the internal consistency 
  of this model can in principle accurately determines $\eta$, and this confirms the contents of section \ref{occore}.
\section{Discussion}\label{dis}
In order to investigated the extra dimension hypothesis by atomic and subatomic objects we construct a quantum mechanics from it's classical counterpart. The considered object assumed to be so massive for two reasons. First for none relativistic demand and second for having enough low quantum length scale. In our building there is only an input functions which is introduce the effects of transition and coming back the quantum test particle in the presence of an extra window in the model. The research about this function guided us to this fact that, it is a quantum wave function for the test particle in to the 2d extra window located at each point of 3d ordinary space. Such wave function is sewn to 3d quantum wave function at the boundary $ \epsilon=1$. Moreover detailed investigation in our constructed quantum mechanic shows that one can derives more information from $\eta(r)$ by a Taylor expansion for it at boundary (locally) and a conjectured duality.
In a diverse viewpoint to constructed quantum mechanics, the attendance to extra dimension in quantum level ultimately converts the Planck constant into the Planck function of 3d space coordinates.
Although we doesn't take special form for $\hbar(r)$ but many results stem from it's general characteristics. For example a candidate for resolving dark matter program is "varying physical constants" in the quantum cosmology. Introducing $\eta(r)$, enters minimally a length scale, inevitably. Such new scale manifested in quantum commutators like Planck constant. For this reason we named the process as double quantization. We suggest that $\rho_{c}$ to be seen by topological (double)quantum effects such as Berry phase, magnetic monopole and internal duality in 5d wave function.
In solving constructed quantum mechanics for free test particle we drive K.K modes effects in the energy spectrum. It was amazing because the investigation was not in a field theory framework, in contrast we were working in a one particle quantum states.


\acknowledgments
Z.GH thanks SKU for their hospitality during the Covid19 era. 



\end{document}